%% file: ring.tex
\documentclass[global]{svjour}
\usepackage{amsmath}
\usepackage{graphicx}
\usepackage{booktabs}

\begin{document}

  \title{The Parametric Transition of Strange Matter Rings to a Black Hole}
  \journalname{General Relativity and Gravitation}
  \author{H.\ Labranche \inst{1}\thanks{\email{H.Labranche@tpi.uni-jena.de}} \and
          D.\ Petroff   \inst{1}\thanks{\email{D.Petroff@tpi.uni-jena.de}}   \and
          M.\ Ansorg    \inst{2}\thanks{\email{mans@aei.mpg.de}}}
          \institute{Theoretisch-Physikalisches Institut, University of Jena, Max-Wien-Platz 1, 07743 Jena, Germany
              \and
                     Max-Planck-Institut f\"ur Gravitationsphysik, Albert-Einstein-Institut, 14476 Golm, Germany}
%
 \maketitle

 \begin{abstract}
  \input abstract
  \keywords{black hole; stationary and axisymmetric spacetimes; rings; numerical solutions;
            strange matter \hfill preprint number: AEI-2006-021}
 \end{abstract}


\section{Introduction}
  \input{intro}

\section{Equation of state}\label{eos}
  \input{eos}

\section{Multipole Moments}\label{multipoles}
  \input{multipoles}

\section{Throat geometry}\label{throat}
  \input{throat}

\section{Escape energy}\label{escape-energy}
  \input{escape-energy}

\section{Summary}\label{summary}
  \input{summary}

 \begin{acknowledgement}
 The authors wish to thank Reinhard Meinel for all his help, Andreas Kleinw\"achter
 for providing us with multipoles for the relativistic disc of dust and Stefan Horatschek
  for many useful discussions.
   This research was funded in part by the Deutsche For\-schungs\-ge\-mein\-schaft
   (SFB/TR7--B1).
 \end{acknowledgement}

 \bibliographystyle{unsrt}
 \bibliography{Reflink} 
  
\end{document}

%% file: abstract.tex
It is shown numerically that strange matter rings permit a continuous
transition to the extreme Kerr black hole. The multipoles as defined by
Geroch and Hansen are studied and suggest a universal behaviour for 
bodies approaching the extreme Kerr solution parametrically. The appearance
of a `throat region', a distinctive
feature of the extreme Kerr spacetime, is observed. With regard to stability,
we verify for a large class of rings, that a particle sitting on the surface of
the ring never has enough energy to escape to infinity along a geodesic.

%% file: intro.tex
In this paper, we consider in detail the parametric
transition of a strange matter ring to a black hole.
In \cite{Meinel04,Meinel06}, necessary
and sufficient conditions for a quasi-stationary
transition were presented and it was proved that an extreme Kerr
black hole necessarily results. Using the analytic
solution for the relativistic disc of dust \cite{NM95}, a
transition to a black hole was found explicitly \cite{Meinel02}.
Transitions have also been found numerically for rings with a variety
of equations of state \cite{AKM4,FHA05}. Such parametric transitions
to a black hole can be interpreted as a quasi-stationary collapse.

The methods we use to study
parametric transitions differ from those in the above cited papers, since we
here concentrate on the behaviour of multipole moments and
on the appearance of a region of spacetime typical of metrics
close to the extreme Kerr limit. These transitions are studied for
strange matter, which is considered to be a form of matter that may
be astrophysically relevant and has not been considered elsewhere
for ring topologies. We restrict our attention here to ring configurations,
since there is evidence suggesting that only rings and discs permit a
transition to a black hole. We include in this paper a comparison with the
corresponding transitions of rings governed by other equations of state.   

Section \ref{eos} is devoted to a brief description of the
equation of state used here to model strange matter.
In Sec.~\ref{multipoles} we define multipole moments at
infinity and follow their progression as they tend to
those of the extreme Kerr black hole. The appearance of
a ``throat region'' separating the ``inner'' from the 
``outer world'' is discussed in Sec.~\ref{throat}. 
In Sec.~\ref{escape-energy}, we verify numerically
that a particle resting on the ring's surface is
always gravitationally bound, a condition, which
can be considered to be a minimal requirement for stability.
We close with a short
summary in Sec.~\ref{summary}.

Throughout this paper, units are used in which the
gravitational constant $G$ and speed of light $c$ are
equal to one.

%% file: eos.tex
Strange matter is a fluid made of up (u), down (d) and strange (s) quarks. Our equation of state (eos) to characterize strange matter is the same one described by Gourgoulhon et al.\ \cite{Gourgoulhon99}, who studied the properties of axially symmetric, stationary, spheroidal strange matter configurations. Based on the MIT bag model, we consider equal numbers of massless, non-interacting u,d,s quarks, confined to a given volume, i.e. enclosed in a ``bag''. The limits of the bag correspond in our case to the surface of the star, such that the star is entirely composed of strange matter. This model leads to a simple eos: 
\begin{equation}
\epsilon = 3p+4B,
\end{equation}
where $\epsilon$ is the energy density, $p$ the pressure and $B$ the bag constant, characterizing the quark confinement.

In the Newtonian limit, the pressure $p$ is low and negligible in comparison to $B$, so the eos takes the form $\epsilon=\text{constant}$. Therefore, all the known Newtonian solutions for homogeneous bodies will be found in the Newtonian limit of the MIT bag model. Of course, a quark model of matter is not relevant in the Newtonian limit, but is taken as a limiting case of our eos. Also, like the homogeneous eos, but unlike polytropic models, the density of strange matter is discontinuous at the surface.

%% file: multipoles.tex
It has been shown in \cite{Meinel04} that the extreme Kerr solution is the only black hole limit of rotating perfect fluid bodies in equilibrium. The extreme Kerr black hole is characterized by the relation
\begin{equation}
J=\pm M^2,
\end{equation}
where $M$ is the mass and $J$ the angular momentum. To study quasi-stationary transitions that lead to black holes, we use bodies with a ring  topology, since spheroidal bodies do not seem to have stationary sequences that
lead to black holes \cite{Ansorgetal04}. For spheroidal bodies, a finite upper bound is observed for $Z_0$, which is the relative redshift of zero angular momentum photons emitted from the surface of the body and observed at infinity. In contrast, the transition to a black hole occurs if and only if $Z_0 \to \infty$ \cite{Meinel06}. We explore here such transitions with the concept of multipole moments.

\subsection{The Metric and the Definition of Multipole Moments}
The line element for a stationary and axisymmetric spacetime containing a uniformly rotating fluid can be written in the form
\begin{equation}
ds^2=e^{-2U}[e^{2k}(d\rho^2+d\zeta^2) +W^2 d\varphi^2]-e^{2U}(a d\varphi +dt)^2, \label{metric}
\end{equation}
where the functions $e^{2k}$, $e^{2U}$, $W$ and $a$ depend only on $\rho$ and $\zeta$. The equatorial
plane is given by $\zeta=0$ and the axis of rotation by $\rho=0$.

To describe the surface of the ring, it is useful to introduce the metric potential $V$,
\begin{equation*}
  e^{2V} = e^{2U} [(1+\Omega a)^2 - \Omega^2 W^2 e^{-4U}],
\end{equation*}
where $\Omega$ is the angular velocity of the fluid with respect to infinity. The function $V$ is constant along isobaric surfaces and the surface of the ring, defined to be the surface of vanishing
pressure, can thus be denoted by $V=V_0$. The constant $V_0$ is related to the relative redshift $Z_0$ via
\begin{equation}
 e^{-V_0} -1 = Z_0.
\end{equation}

Consider now the vacuum region exterior to the mass distribution and extending to infinity.
It is not possible to solve Einstein's equations in this region alone since the boundary
conditions valid on the surface of the ring (and indeed the location of this surface) can
only be found after solving the global problem. Considering the vacuum region
will suffice for introducing the multipole moments however. In this region,
there exists a conformal coordinate transformation $z'=z'(z)$ ($z':=\rho'+ i \zeta'$, $z:=\rho+ i \zeta$)
allowing one to choose $\rho'(\rho,\zeta)=W(\rho,\zeta)$, which then leads to the metric
\begin{equation}
ds^2=e^{-2U}[e^{2k'}(d\rho'^2+d\zeta'^2) +\rho'^2 d\varphi^2]-e^{2U}(a d\varphi +dt)^2, \label{metric2}
\end{equation}
which we will use below to define the multipole moments. Note that the Cauchy-Riemann conditions
for the transformation from (\ref{metric}) to (\ref{metric2})
imply $W,_{\rho\rho}+W,_{\zeta\zeta}=0$, which is valid only in the vacuum domain. It follows from
axial symmetry that $W=0$ holds for $\rho=0$ ($W=\cal{O}(\rho)$).
Along the axis of rotation, $\rho=0$, one of the Cauchy-Riemann conditions
then yields
\begin{align}
 \left.\frac{\partial\zeta'}{\partial\zeta}\right|_{\rho=0} =&\, 
 \left.\frac{\partial\rho'}{\partial\rho}\right|_{\rho=0}
         = \left.\frac{\partial W}{\partial\rho}\right|_{\rho=0} 
         =\, \lim_{\rho \to 0}\frac{W}{\rho}.
\end{align}
After solving numerically for the metric functions in Eq.~(\ref{metric}), an expansion of $W/\rho$
along the axis of rotation then allows us to find an expansion of $\zeta'$ in terms of $\zeta$
and vice versa. Thus, taking into account that $\rho=0 \Leftrightarrow \rho'=0$, we are in a
position to be able to write down the series expansion about the point $\zeta'=+\infty$ for
the metric functions,
which will be used in defining the multipole moments (see Eq.~\ref{xi_series}).

Turning our attention back to Eq.~(\ref{metric2}), the Einstein equations governing $a$ and $e^{2U}$ can be rewritten using the single, complex Ernst equation
\begin{equation}
(\Re f) \triangle f = \nabla f \cdot \nabla f,
\end{equation}
where $f$ is the complex function $f=e^{2U}+ib$, $\Re f$ is the real part of $f$ and $\triangle$ and $\nabla$ are respectively the Laplacian and the gradient operators in a three dimensional Euclidean space. Once $a$ and $U$ have been solved for, the metric function $k'$ can be calculated via a line integral. Solutions of the Ernst equation lead to solutions of the Einstein equations and the metric potentials can be calculated from:
\begin{eqnarray}
a,_{\rho'}  & = &  \rho' e^{-4U} b,_{\zeta'} \\
a,_{\zeta'} & = & -\rho' e^{-4U} b,_{\rho'} \\
k',_{\rho'}  & = &  \rho' [U,_{\rho'}^2-U,_{\zeta'}^2+ \frac{e^{-4U}}{4} (b,_{\rho'}^2-b,_{\zeta'}^2)] \\
k',_{\zeta'} & = & 2\rho' [U,_{\rho'} U,_{\zeta'}+\frac{e^{-4U}}{4} b,_{\rho'} b,_{\zeta'}].
\end{eqnarray}
From the Ernst potential $f$, one can define another potential $\xi$:
\begin{equation}
\xi = \frac{1-f}{1+f}.
\end{equation}
Taking the potential $\xi$ on the positive part of the axis of rotation ($\rho'=0$, $\zeta'>0$), we can make a series expansion of it at infinity:
\begin{equation}\label{xi_series}
\xi(\rho'=0,\zeta') = \sum_{n=0}^{\infty} \frac{m_n}{\zeta'^{n+1}}.
\end{equation}
We assume reflectional symmetry about the equatorial plane in this paper
for which it follows that $m_n$ is real for even $n$ and imaginary
for odd $n$ \cite{Kordas95,MN95}.

The multipole moments $P_n$ defined by Geroch \cite{Geroch70} and Hansen \cite{Hansen73}
are algebraic combinations of the coefficients $m_n$ and characterize the Ernst potential
uniquely. In this article, we consider the 7 first
multipole moments of the infinite set given in \cite{Fodoretal89} as
\begin{subequations}
\begin{align}
P_j  =~& m_j \quad \text{for} ~ j=0,1,2,3   \\
P_4  =~& m_4 -\frac{1}{7}M_{20} m_0   \\
P_5  =~& m_5 -\frac{1}{3}M_{30} m_0 + \frac{1}{21}M_{20} m_1\\
\begin{split}
P_6  =~& m_6 +\frac{1}{33}M_{20} m_0^3 -\frac{5}{231}M_{20} m_2 \\ & +\frac{4}{33}M_{30} m_1  
        -\frac{8}{33}M_{31} m_0  -\frac{6}{11}M_{40} m_0, 
\end{split}
\end{align}
\end{subequations}
where $M_{jk} \equiv m_j m_k - m_{j-1} m_{k+1}$. We point out that $P_0=M$ and $P_1=iJ$
always hold. The multipoles $P_n$ can then be normalized as follows:
\begin{equation}
y_n = i(-2i\Omega)^{n+1} P_n.
\end{equation}

For the Kerr black hole, the multipole moments are simply
\begin{equation}\label{P_n}
	P_n^{(\text Kerr)} = M (iJ/M)^n,
\end{equation}
where $M$ and $J$ are respectively the mass and the angular momentum of the black hole.
Using the relation
\begin{equation}
	J=\frac{4\Omega_{\text H} M^3}{1 + 4\Omega_{\text H}^2 M^2},
\end{equation}
where $\Omega_{\text H}$ is the angular velocity of the horizon, we then find
\begin{equation}\label{yn_for_Kerr}
	y_n^{(\text Kerr)}(y_0) = y_0\left(\frac{2y_0^2}{1+y_0^2}\right)^n.
\end{equation}
Through this normalization, all multipoles $y_n$ of the extreme Kerr black hole
are equal to one, as can be seen by taking into account $y_0=2\Omega_{\text H} M=1$.

\subsection{Multipole Moments of Rings}
As $V_0$ tends to $-\infty$, we expect the multipole moments to become closer and
closer to those of an extreme Kerr black hole. We tested this numerically by making use
of a (slightly modified version of a) highly accurate computer program as described in
\cite{AKM3}. This program was used for all the results presented in this paper.
Since the $n^{\text{th}}$ multipole moment requires the calculation of $n+1$
derivatives, which results in a loss of accuracy, we could not calculate arbitrarily many
numerically.

\begin{figure}
 \centerline{\includegraphics{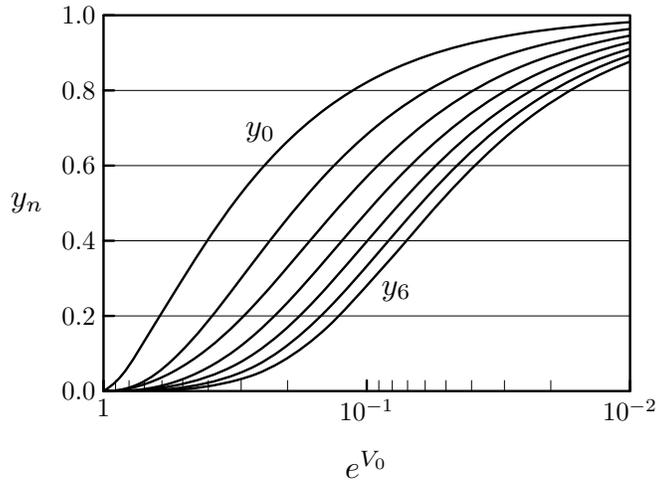}}
 \caption{The normalized multipoles $y_n$ 
 versus $e^{V_0}$ for homogeneous rings
 with $\rho_\text i/\rho_\text o=0.7$ . \label{multihom}}
\end{figure}

\begin{figure}
 \centerline{\includegraphics{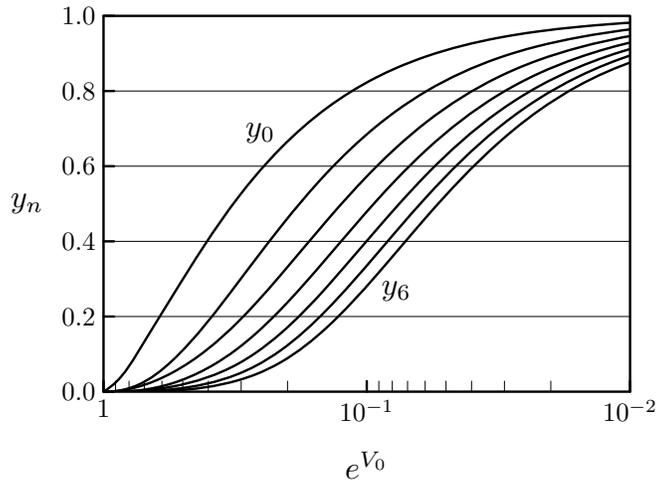}}
 \caption{The normalized multipoles $y_n$  versus $e^{V_0}$ for strange matter rings
 with $\rho_\text i/\rho_\text o=0.7$. \label{multistr}}
\end{figure}

\begin{figure}
 \centerline{\includegraphics{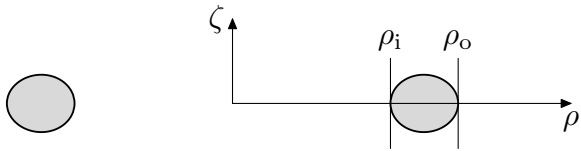}}
 \caption{Example of a meridional cross-section of a strange matter ring.  The ring in this example has the parameters 
          $\rho_\text i/\rho_\text o=0.7$ and $e^{2V_0}=0.1$.\label{ring}}
\end{figure}

Figures~\ref{multihom} and \ref{multistr} show the first seven multipole moments for
homogeneous and strange matter rings where the ratio between the inner coordinate radius
$\rho_\text i $ and the outer radius $\rho_\text o$ (see Fig.~\ref{ring}) is held constant
at a value of $\rho_\text i/\rho_\text o=0.7$. The left side of the plots corresponds
to the Newtonian limit and the right side tends toward the black hole
limit. As $V_0 \to -\infty$, the normalized multipoles all tend to one,
demonstrating that this sequence indeed approaches the extreme Kerr
solution.

It is interesting to note, with respect to $e^{V_0}$ (or $Z_0$),
how slowly the exterior spacetime approaches
that of a Kerr black hole. Consider, for example, the configuration from
Fig.~\ref{multistr} with $e^{V_0}=10^{-2}$. Whereas the value
$J/M^2=1.00014$ is very close to the limiting value of one reached in
the extreme Kerr limit, the product
$2\Omega M=0.9813$ deviates rather significantly from it. This makes
itself felt particularly for the higher multipole moments where
powers of $\Omega$ are in play. The moment $y_4$, for example, has
reached only a value of $y_4 \approx 0.91$ for this configuration.

\begin{figure}
 \centerline{\includegraphics{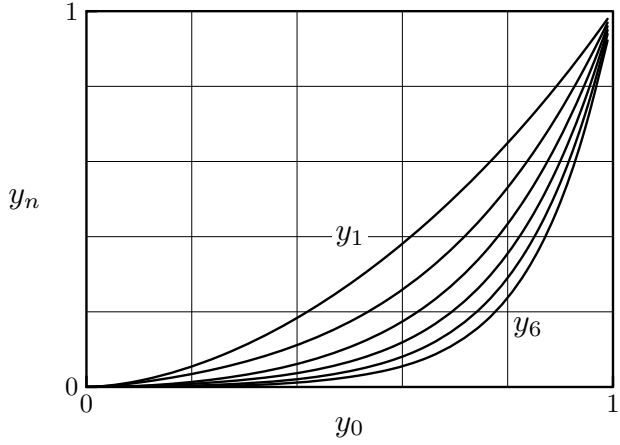}}
 \caption{The multipole moments $y_n$, $n=1 \ldots 6$ versus $y_0$
      for strange matter rings with $\rho_\text i/\rho_\text o=0.7$. \label{y_n_ring}}
\end{figure}

\begin{figure}
 \centerline{\includegraphics{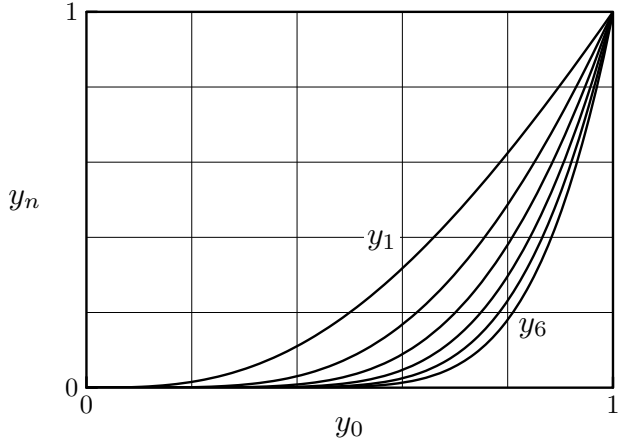}}
 \caption{The multipole moments $y_n$, $n=1 \ldots 6$ versus $y_0$
      for the sequence of Kerr solutions.\label{y_n_Kerr}}
\end{figure}

\begin{figure*}
\centerline{\includegraphics{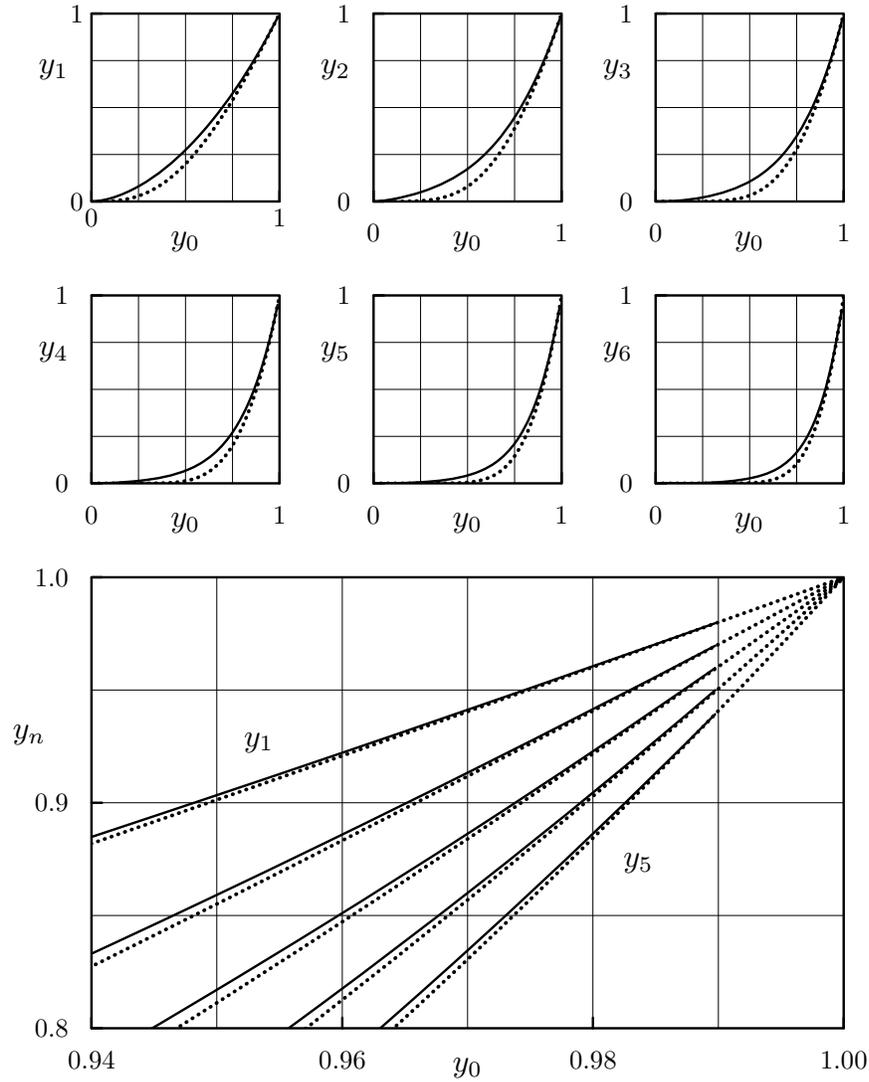}}
 \caption{Various multipole moments $y_n$ are plotted versus $y_0$ 
    for strange matter rings with $\rho_\text i/\rho_\text o=0.7$ (solid lines)
	 and the sequence of Kerr solutions (dotted lines). In the detailed
	 plot, the curve for $y_6$ was omitted because of slight numerical
	 inaccuracies for higher multipole moments. \label{y_n}}
\end{figure*}

To understand better the nature of the transition to the black hole, we
compare the multipole moments of the above strange matter ring sequence
with those of the Kerr solution. 
In Fig.~\ref{y_n_ring} the $y_n$ for $n=1\ldots 6$ are plotted against $y_0=2 \Omega M$
for the strange matter ring sequence from above. A
corresponding picture for the sequence  of Kerr solutions (see (\ref{yn_for_Kerr})) is displayed in
Fig.~\ref{y_n_Kerr}. The clear similarity between  these plots is emphasized in Fig.~\ref{y_n}
where each $y_n$ for the ring (solid line) and the Kerr solution (dotted line) is compared
in a small figure over its whole range. The region very close to the extreme Kerr limit is
then shown for $y_1$--$y_5$ in detail. The graphs strongly suggest that the slopes
\begin{equation}\label{slope_of_yn_1}
	\frac{dy_n}{dy_0}(y_0=1)
\end{equation}
are the same for the Kerr family and for the strange matter ring sequence discussed here. In fact, we found 
these slopes to be indepedent of the specific eos being used.%
\footnote{We checked this for ring sequences 
governed by homogeneous, polytropic and Chandrasekhar 
eos as well as for the rigidly rotating dust family.
The Chandrasekhar eos describes a completely degenerate, zero temperature,
 relativistic Fermi gas.}
For the Kerr solutions, it follows from (\ref{yn_for_Kerr}) that
\begin{equation}\label{slope_of_yn_2}
	\frac{dy_n}{dy_0}(y_0=1)=n+1,
\end{equation}
which leads us to the conjecture that (\ref{slope_of_yn_2}) holds true for all sequences of rotating 
bodies that admit the transition to an extreme Kerr black hole. This conjecture
provides a universal growth rate with which the $y_n$ approach unity. It would be interesting to
test this conjecture for the analytically known rigidly rotating disc of dust and work in this
direction is ongoing.

In Table~\ref{tab:moments_Z0=99}, a comparison of the values of the first
five moments $y_n$ for a variety of configurations all with $e^{V_0}=10^{-2}$ is provided.
The set of configurations chosen includes rings with various different eos
 and various radius ratios and also includes the uniformly rotating disc of dust.
A discussion of the multipoles of this last configuration as well as plots analogous
to Fig.~\ref{multistr} can be found in \cite{KMN95}. Since all multipole moments tend to
one in the limit $V_0 \to -\infty$, these multipoles will provide almost no way
of distinguishing between various configurations close to this limit.

\begin{table}
 \centerline{\begin{tabular}{cccccc}
  \toprule
   & $y_0$ & $y_1$ & $y_2$ & $y_3$ & $y_4$ \\ \midrule 
  s.m. ($r_\text i/r_\text o=0.6$) & 0.982 & 0.964 & 0.947 & 0.930 & 0.913 \\
  s.m. ($r_\text i/r_\text o=0.7$) & 0.981 & 0.963 & 0.945 & 0.928 & 0.910 \\
  s.m. ($r_\text i/r_\text o=0.8$) & 0.981 & 0.962 & 0.943 & 0.925 & 0.907 \\
  hom. ($r_\text i/r_\text o=0.7$) & 0.981 & 0.963 & 0.945 & 0.927 & 0.910 \\
  pol. ($r_\text i/r_\text o=0.7$) & 0.982 & 0.965 & 0.948 & 0.931 & 0.914 \\
  rel. disc of dust                & 0.984 & 0.969 & 0.953 & 0.938 & 0.924 \\
 \bottomrule
  \end{tabular}}
 \caption{The multipole moments $y_n$ for various configurations, all
   with $e^{V_0}=10^{-2}$. The abbreviation `s.m.' refers to a strange matter ring,
   `hom.' to a homogeneous ring, `pol.' to a polytropic ring with the
   polytropic index $n=1$ and `rel. disc of dust' to the relativistic
   disc of dust. \label{tab:moments_Z0=99}}
\end{table} 

In contrast, we present the multipole moments for configurations near the
Newtonian limit ($e^{-V_0}=1.1$) in Table~\ref{tab:moments_Z0=0.1}. Here one can
see that there is far more variation amongst the rings and that the disc
of dust differs significantly from any of the rings.
The values in the table also reflect
the fact that strange matter has the same Newtonian limit as homogeneous matter.

\begin{table*}
 \centerline{\begin{tabular}{cccccccc}
   \toprule
    & $y_0$ & $y_1$ & $y_2$ & $y_3$ & $y_4$  \\
    & $(\times 10^{-2})$ & $(\times 10^{-3})$
    & $(\times 10^{-3})$ & $(\times 10^{-5})$
    & $(\times 10^{-5})$ \\ \midrule
  s.m. ($r_\text i/r_\text o=0.6$) & 2.22 & 1.21 & 1.04  & 8.92 & 7.44\\
  s.m. ($r_\text i/r_\text o=0.7$) & 2.09 & 1.16 & 1.02  & 8.69 & 7.43\\
  s.m. ($r_\text i/r_\text o=0.8$) & 1.92 & 1.09 & 0.978 & 8.49 & 7.46\\
  hom. ($r_\text i/r_\text o=0.7$) & 2.09 & 1.16 & 1.01  & 8.68 & 7.42\\
  pol. ($r_\text i/r_\text o=0.7$) & 2.14 & 1.27 & 1.11  & 10.1 & 8.75\\
  rel. disc of dust                & 2.36 & 1.73 & 1.56  & 12.5 & 21.7 \\
  \bottomrule
  \end{tabular}}
 \caption{The multipole moments $y_n$ for various configurations, all
   with $e^{-V_0}=1.1 \Leftrightarrow Z_0=0.1$. The configurations are labelled as in
   Table~\ref{tab:moments_Z0=99}.
   \label{tab:moments_Z0=0.1}}
\end{table*}

\begin{figure*}
  \centerline{\includegraphics{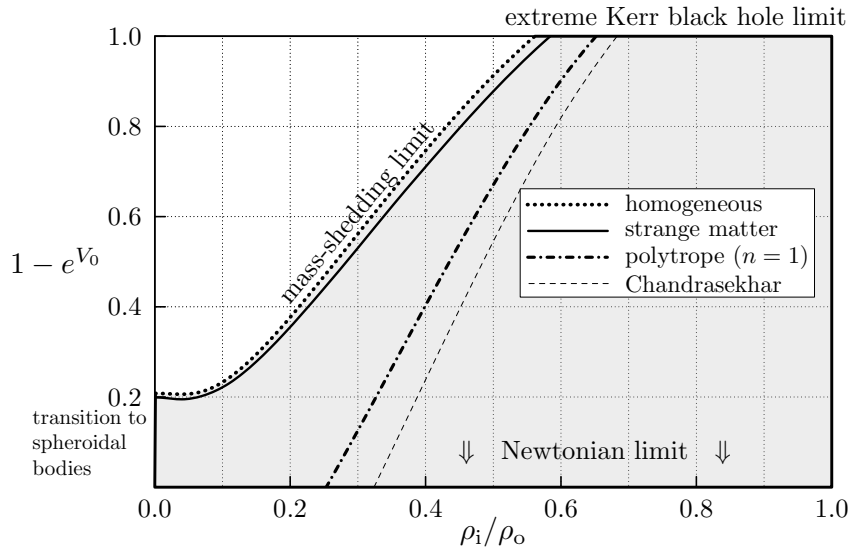}}
 \caption{The parameter space for rings with a variety of eos
          is considered in the ($\rho_\text i/\rho_\text o$)--($1-e^{V_0}$) plane.
          Each eos is bounded on the left by the corresponding mass-shedding
          curve. \label{mass_shed}}
\end{figure*}

A further comparison of rings of various eos can be found in
Fig.~\ref{mass_shed}. Sequences of rings rotating at the mass-shedding limit,
are plotted in a two-dimensional parameter space with $1-e^{V_0}$ on the $y$-axis
and $\rho_\text i/\rho_\text o$ on the $x$-axis. The mass-shedding limit is reached
when the path followed by a particle rotating at the outer
edge of the ring becomes a geodesic. For a given eos, other ring
configurations (i.e.\ not rotating at the mass-shedding limit) lie to the right
of the corresponding curve. One can see that a transition to the extreme Kerr
black hole is a generic feature of all rings considered here. The
transition to spheroidal bodies exists for strange matter rings, but not for all
eos. What is particularly striking is how close together the
curves for strange and homogeneous rings remain right up to the black hole limit.
This figure is a modified version of Fig.~1 of \cite{FHA05}. A discussion of the
polytropic and Chandrasekhar eos can also be found in that paper. 

%% file: throat.tex
One of the most interesting features of bodies near the extreme Kerr
black hole limit is the appearance of a throat geometry
\cite{BH99,Meinel02}. In the limit, the throat separates the `inner world',
containing the ring, from the `outer world'. The outer world is the asymptotically
flat extreme Kerr spacetime, which is described by a single parameter
and in which the horizon is located
at the end of the infinitely long throat. On the other hand, the inner world is not
asymptotically flat and is related to the outer world through its
asymptotic behaviour, which contains information about the one free parameter
that uniquely describes the outer world.  Any point in the outside 
world is infinitely far away from any point in the inner world. For example,
in the equatorial plane, one finds that the radial proper distance $\delta$
from the point $\rho=0$ to the point $\rho$ is
\begin{equation}
  \delta=\int_{0}^{\rho} \left.\sqrt{g_{\rho\rho}}\right|_{\zeta=0}\,\, d\rho.
\end{equation}
If one measures the proper distance from a point $\rho_1$ to a point $\rho$
in the above equation, then it tends logarithmically to $\infty$ for the
extreme Kerr black hole as $\rho_1\to 0$ (the horizon in the coordinates used here
is located at $\rho=0$)%
\footnote{It is thus to be expected that as $Z_0$ becomes large, the
   outer coordinate radius of the ring, as measured in units
   of mass for example, will become small. The shape of the
   ring itself does not change qualitatively as can be seen in
   Fig.~3 of \cite{AKM4} which depicts the shape of the ring along a
   sequence tending to the black hole limit.}.

\begin{figure*}
 \includegraphics{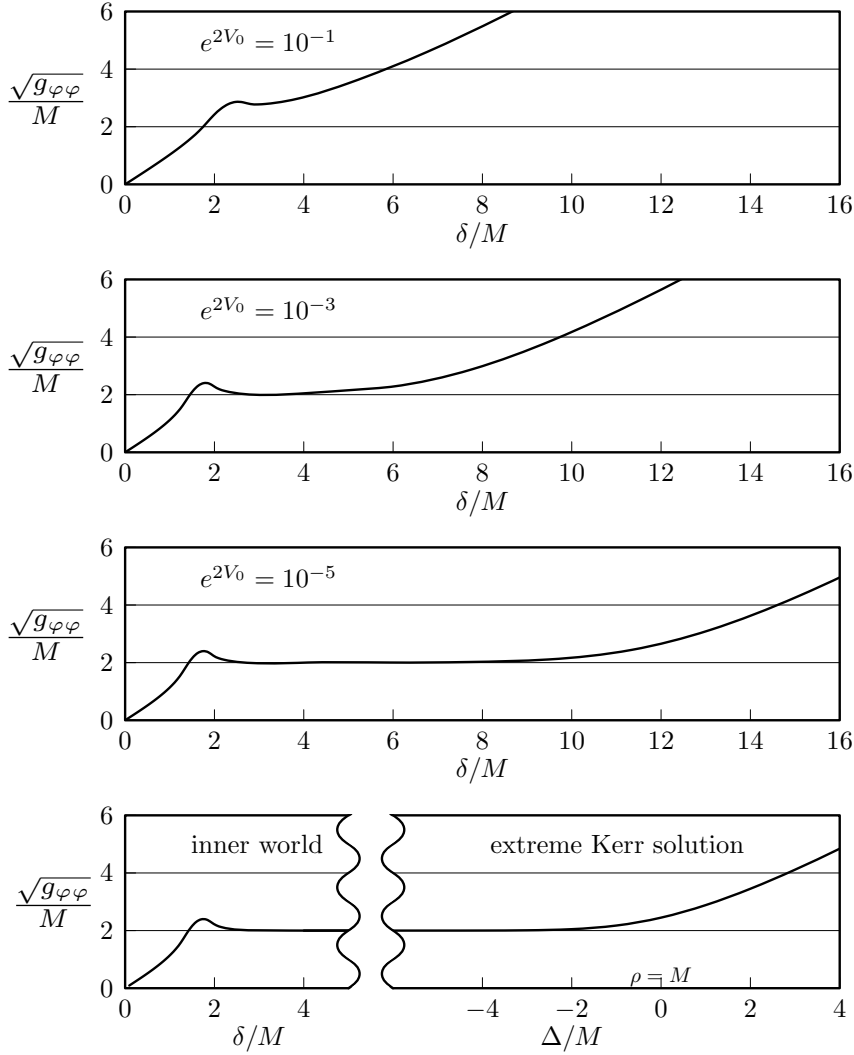}
 
 \caption{The function $\sqrt{g_{\varphi\varphi}}$ in the equatorial plane is plotted
          versus proper distance, both normalized with respect to the mass $M$.
          In the throat region, $\sqrt{g_{\varphi\varphi}}/M$ tends to the constant
          value 2.
          All four plots were made for a strange matter ring with a radius
          ratio $\rho_\text i/\rho_\text o=0.7$ and with a value for $e^{2V_0}$
          as indicated. In the last plot, $\Delta$ gives the proper distance in
          the Kerr metric to the reference point $\rho=M$. Note that the proper
          distance between any point in the `inner world' region and any point
          in the `extreme Kerr' region tends to infinity as $e^{2V_0}\to 0$.
          \label{fig_throat}}
\end{figure*}

One way to represent the throat is to plot $\sqrt{g_{\varphi\varphi}}/M$
in the equatorial plane as a function of $\delta/M$. Then, the throat
appears as a plateau, i.e.\ a region appears in which the circumference
of a circle of constant radius $\rho=\rho_\text c$, tends toward a constant,
independent of the radius $\rho_\text c$. As the extreme Kerr black hole
is approached, this region becomes infinitely long. Figure~\ref{fig_throat}
shows the appearance of the throat for a sequence of strange matter rings
with $\rho_\text i/\rho_\text o=0.7$ as the parameter $e^{V_0}$ tends to zero.
Even in the first of these pictures
($e^{2V_0}=10^{-1}$), the highly relativistic
nature of the ring is demonstrated by the fact that a small portion of the
curve has a negative slope. That is, there exists a region of spacetime in
which circles lying in the equatorial plane and centred about the origin
have decreasing circumference with increasing radius. The last
of these pictures is similar to Fig.~13 in \cite{BW71} in which the `inner world'
is separated from the extreme Kerr solution by the infinite throat region.
The proper distance between a point in what becomes the inner world (e.g.\
the outer edge of the ring $\rho=\rho_\text o$) and a point in what becomes
the outer world (e.g.\ $\rho=M$) tends to infinity as $e^{2V_0}\to 0$.
In a sense, we can say that the `throat region' near the black hole limit `swallows'
the information as to what kind of configuration is sitting at the centre,
as can be seen in Table~\ref{tab:moments_Z0=99}.

The numerical `inner world' solution was produced with a program that prescribes
the asymptotic behaviour of the throat region (see \cite{BH99}). Since the
`asymptotically flat computer program' is capable of rendering rings with a relative
redshift $Z_0$ well in excess of 100, the metric behaviour provided by this
program can be used as initial input for the Newton-Raphson method of
the `inner world program' \cite{AKM3}.  The fact that such inital data converges
to an inner world solution is in itself strong numerical evidence for the existence
of a continuous transition to the black hole. 

%% file: escape-energy.tex
With the four-velocity $u^i$ and the Killing vector $\xi = \partial/\partial t $ corresponding
to stationarity, one can define the specific energy of a test particle with respect to
infinity, i.e.\ the energy per unit mass, as
\begin{equation}
E = -u^i \xi_i \label{escape-E},
\end{equation}
which is a conserved quantity along any geodesic. 

With $u^i$ referring to the four-velocity of a particle resting on the ring's surface,
$E-1$ could be called the ``escape energy''. If it is negative, then a sufficiently small
perturbation will not suffice to induce the particle's escape to infinity on a geodesic,
and it is referred to as gravitationally bound. In proving that $V_0 \to -\infty$ is a
sufficient condition for reaching the Black Hole limit \cite{Meinel06}, use was made
of the reasonable assumption that particles on the fluid's surface are gravitationally
bound. One expects that this minimal requirement for stability will always be satisfied.
We now proceed to verify this assumption for a large class of rings.  

\begin{figure}
 \centerline{\includegraphics{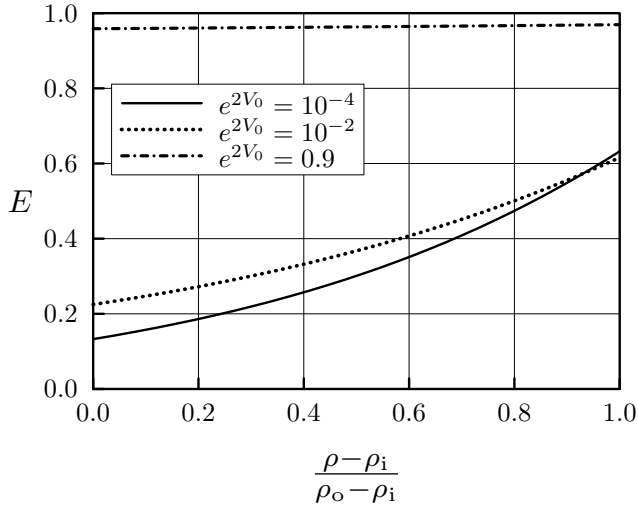}}
 \caption{The specific energy $E$ versus $(\rho-\rho_\text i)/(\rho_\text o-\rho_\text i)$
          on the surface of a variety of strange matter rings with $\rho_\text i/\rho_\text o=0.7$.
          \label{en-variable-rho-str_a}}
\end{figure}

\begin{figure}
 \centerline{\includegraphics{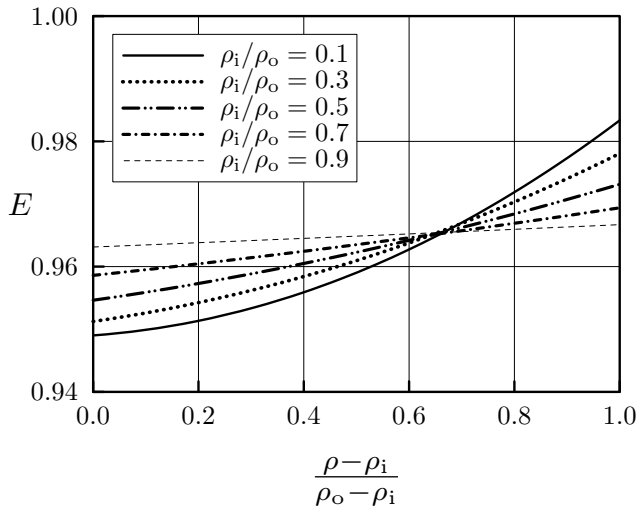}}
 \caption{The specific energy $E$ versus $(\rho-\rho_\text i)/(\rho_\text o-\rho_\text i)$
          on the surface of a variety of strange matter rings near the Newtonian limit ($e^{2V_0}=0.9$).
          \label{en-variable-rho-str_b}}
\end{figure}

Figures~\ref{en-variable-rho-str_a} and \ref{en-variable-rho-str_b} show
the value of $E$ along the surface of a variety of strange
matter rings as it depends on radius.
The radial parameter $(\rho-\rho_\text i)/(\rho_\text o-\rho_\text i)$
is chosen such that it runs from 0 to 1 for every ring. In Fig.~\ref{en-variable-rho-str_a} curves are plotted for a constant value $\rho_\text i / \rho_\text o =0.7$ and for varying $V_0$. We see that $E$ tends to 1 in the Newtonian limit, which follows directly from Eq.~(\ref{escape-E}). Figure~\ref{en-variable-rho-str_b} shows the behaviour of $E$ for various values of $\rho_\text i / \rho_\text o$ and constant $V_0$. Since configurations with small $\rho_\text i / \rho_\text o$ do not exist when $V_0$ becomes too negative (see Fig.~\ref{mass_shed}), we chose $V_0$ to be in the Newtonian regime in order to be able to consider a wide range of values for the radius ratio. For every example considered in Figs~\ref{en-variable-rho-str_a} and \ref{en-variable-rho-str_b}, a maximal value at the outside edge of
the ring in the equatorial plane is reached, just as one would expect. 
It is interesting to compare these results with the relativistic
disc of dust for which $E=1$ holds at the outer edge independent
of the value of $Z_0$ \cite{MK95}.

\begin{figure}
 \centerline{\includegraphics{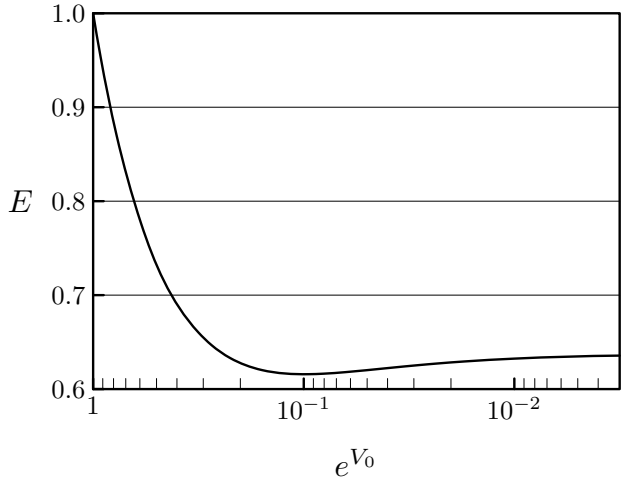}}
 \caption{The specific energy $E$ versus $e^{V_0}$ at the outer edge in the equatorial plane
          for strange matter rings with $\rho_\text i/\rho_\text o=0.7$. \label{en-str}}
\end{figure}

Focussing our attention now on the outer edge of the ring in the
equatorial plane, we see in
Fig.~\ref{en-str} how $E$ depends on $V_0$ for
a sequence of strange matter rings with $\rho_i/\rho_o=0.7$.
It is apparent that a maximum is reached in the Newtonian
limit. For rings rotating at the mass-shedding limit, the
value of $E$ is also significantly smaller than one for
small $e^{V_0}$. The results for homogeneous rings are very similar
and we can verify that $E \le 1$ holds (i.e. the escape energy is negative) 
for a large class of rings. 

%% file: summary.tex
It was shown numerically that a parametric transition exists from
strange matter rings to the extreme Kerr black hole. Whereas it is
known analytically that the eos describing strange matter tends to
that of a homogeneous body in the Newtonian limit, it was shown here
that properties of configurations with these two eos
remain similar right up to the limit $V_0 \to -\infty$.

Figs~\ref{multihom} and \ref{multistr} suggest that the transition to
the black hole is rather slow as $e^{2V_0} \to 0$. This can be
made more precise through the comparison with the Kerr solution,
which leads us to conjecture that for every stationary rotating body
permitting a transition to a black hole, the multipole moments $y_n$
tend to one according to the formula
\[ \frac{dy_n}{dy_0}(y_0=1)=n+1.\]

It is expected that $E \le 1$ always holds on the surface of a
fluid body. Indeed, this inequality was used to prove that for rotating
fluids the extreme Kerr black hole necessarily results if $e^{2V_0} \to 0$
\cite{Meinel06}. We have verified that this
inequality is correct for a large class of rings.

As our knowledge of astrophysical collapse scenarios improves,
it will be interesting to see how strong the connections can
be to the quasi-stationary collapse considered here.